\documentclass{ws-procs961x669}            

\def\be{\begin{equation}}
\def\bea{\begin{eqnarray}}
\def\eea{\end{eqnarray}}
\def\ee{\end{equation}}
\begin{document}
\title{GRB prompt phase spectra under backscattering dominated model}

\author{Mukesh Kumar Vyas$^*$ and Asaf Pe'er}

\address{Department of Physics, Bar Ilan University,,\\
Ramat Gan, 5290002, Israel\\
$^*$E-mail: mukeshkvys@gmail.com\\
asaf.peer@biu.ac.il\\
www.biu.ac.il}

\author{David Eichler}
\address{Ben-Gurion University,\\
Be'er Sheva, 84105, Israel\\
}

\begin{abstract}
We propose a backscattering dominated prompt emission model for gamma ray bursts (GRB) prompt phase in which the photons generated through pair annihilation at the centre of the burst are backscattered through Compton scattering by an outflowing stellar cork. We show that the obtained spectra are capable of explaining the low and high energy slopes as well as the distribution of spectral peak energies in their observed prompt spectra.
\end{abstract}


\bodymatter

\section{Introduction}
In the picture of gamma ray bursts (GRBs), collapse of a massive star gives rise to a long GRB, while a merger of two compact objects leads to a short GRB. In both the cases, strong bipolar jets are produced \citep{levinson1993baryon, macfadyen1999collapsars}. Radiation from these jets is observed if the observer is near the jet axis. The initial phase of the burst is known as prompt phase. In a typical understanding of this phase, the jet propagates inside the stellar interior and forms an envelope (stellar cork). Finally, the jet breaks out of the cork to be observed. \citep{ramirez2002precursors, zhang2003relativistic} 

An alternate view of this stage was proposed, \cite{eichler2008spectral, eichler2014cloaked, eichler2018short, vyas2021backscattering, vyas2021predictingAPJL} in which the core of the star produces electron-positron pair plasma through neutrino annihilation. This pair plasma emits radiation in form of a jet through pair annihilation and pushes the stellar material to form the stellar cork. This cork is further pushed to relativistic speeds by radiation pressure. The radiation is not able to pierce the cork and is backscattered. As the cork is moving with relativistic speeds, this radiation is beamed along the direction of the jet and is seen by the observer.

In this work, we carry out Monte Carlo simulations of the interaction of the seed photons with the cork where the backscattered photons lead to the observed spectrum. We show that the spectrum observed is capable of explaining a large range of observed spectral properties like the photon indices at low and high energies along with the spectral peak energies.
\begin{figure}
\begin{center}
\includegraphics[width=2.5in]{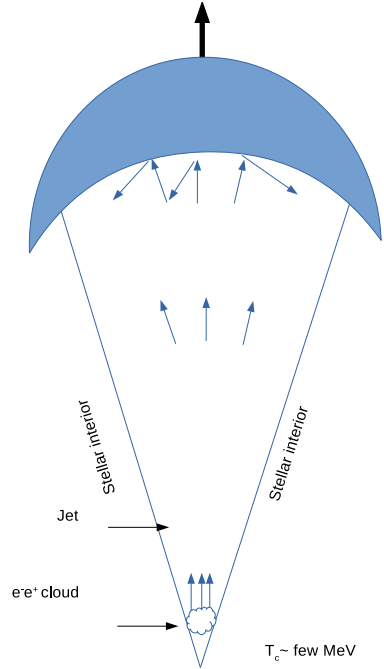}
\caption{Geometric representation of the system. The seed photons are generated near the centre of the bursts through pair annihilation and comptonized with the plasma present there. These photons propogate in empty jet funnel and interact with a radially outflowing stellar cork.}
\label{aba:fig1}
\end{center}
\end{figure}
\section{Physical picture of the system and source of the seed photons}
At the time of stellar collapse, neutrinos are produced and get annihilated near the centre of the star. This annihilation produces an electron-positron pair dominated plasma. This pair plasma produces annihilation spectrum along with bremmstrahlung and is further Comptonized by the local plasma. Using Monte Carlo simulations, the resultant spectra were studied by \cite{zdziarski1984effect}. The spectra were found to be flat with exponential decay at high energies. We obtained numerical fits to his results. For plasma density $n=2\times 10^{18}$ cm$^{-3}$ the fitted spectrum is 
\be 
F_{\varepsilon}=C_0 \exp \left(-\frac{C_1 \varepsilon^2}{\Theta_r^2}\right) {~~~\rm KeV/s/}{\rm KeV}^{-1}
\label{eq_zd_fit}
\ee
Here $\varepsilon$ photon energy normalized to electron's rest mass energy. $\Theta_r=k_BT_r/m_ec^2$ is dimensionless representation of pair temperature $T_r$ with $k_B$, $m_e$ and $c$ are Boltzmann's constant, rest mass of the electron and light speed, respectively. $C_1=0.045$ and $C_0(=2\times 10^{40}$ KeV s$^{-1}/$KeV). This fit gives reliable spectra for temperatures $\Theta_r>0.3$.
The photons with this spectral shape propagate within the jet funnel and push the stellar matter that forms a stellar cork ahead of it. The cork moves with relativistic speeds and escapes the stellar surface. The seed photons enter the hot cork with temperature $T_c$ (or $\Theta_c=k_BT_c/m_ec^2$). After going through Compton scattering with the electrons inside the cork, the photons gain or loose energy and get backscattered from the rear end. The geometry of this figure is explained in Figure \ref{aba:fig1}. Because of the relativistic motion of the cork with Lorentz factor $\gamma$, these photons are then observed by an observer situated at angle $\theta_{obs}$ from the jet axis. The cork has curved inner surface and the photons scattered at different locations are observed at different times as well as with different energies producing a light curve as well as spectrum. 

\section{Solution procedure}
Through a Monte Carlo code, we inject approximately $ 26$ million seed photons with the spectral distribution according to Equation \ref{eq_zd_fit}. These photons are launched at an adiabatically expanding and outflowing cork with temperature $T_c$ and Lorentz factor $\gamma$ with an opening angle $\theta_j=0.1$ rad. The photons enter the cork from its rear end (Figure \ref{aba:fig1}) and scattered backwards. Before escaping, the photons go though multiple scattering with energetic electrons inside the cork and thus their energies evolve. More details of the calculation procedure of the Monte Carlo code is described before. \cite{pe2008temporal, vyas2021backscattering} These photons are then observed in specific angular patch $d\theta_{obs}=0.005$ rad about the observer's location $\theta_{obs}$. The resultant spectrum is shown in the rest frame of the GRBs with an assumption that the total energy released from the centre is equivalent to $10^{50}$ erg.
\section{Results}
\begin{figure}
\begin{center}
\includegraphics[width=2.5in]{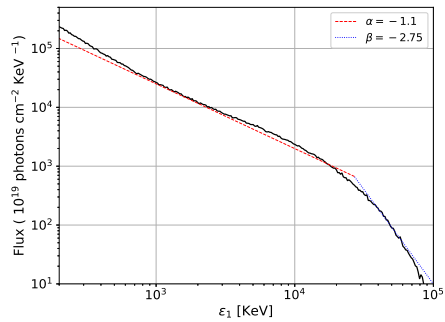}
\caption{Photon spectrum obtained for $\Theta_c=1.4$, and $\theta_j=0.1$rad. The observer is situated at $\theta_{obs}=0.05$ rad. Here $\Theta_r=3.0$ and $\gamma=100$}
\label{aba:fig2}
\end{center}
\end{figure}

\begin{figure}
\begin{center}
\includegraphics[width=2.5in]{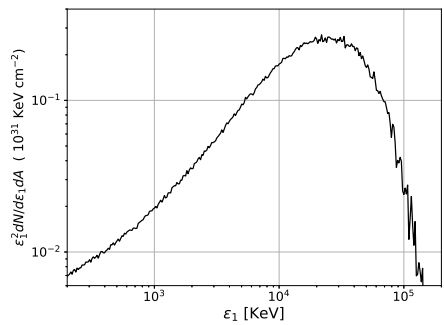}
\caption{Emissivity KeV cm$^{-2}$ for the same parameters as in Figure \ref{aba:fig2}}
\label{aba:fig3}
\end{center}
\end{figure}

In Figure \ref{aba:fig2}, we plot a typical photon spectrum (photons cm$^{-2}$KeV $^{-1}$) obtained by scattering of seed photons with $\Theta_r=3$ and cork temperature $\Theta_c=1.4$. The cork has a bulk Lorentz factor $\gamma=100$ and initial location at $r_i=3\times 10^{12}$cm from the centre of the star. The opening angle of the cork is taken to be $\theta_j=0.1$ rad. The spectrum obtained has a low energy photon index $\alpha=-1.1$ while high energy index is $\beta=-2.75$. In Figure \ref{aba:fig3} we show the variation of spectral emissivity KeV cm$^{-2}$. The Spectral peak is obtained to be at $\varepsilon_{peak}=30$ MeV. This typical spectrum obtained resembles the well known GRB prompt phase spectral shape with two power laws separated by a spectral peak. \cite{preece2000batse}

\begin{figure}
\begin{center}
\includegraphics[width=2.5in]{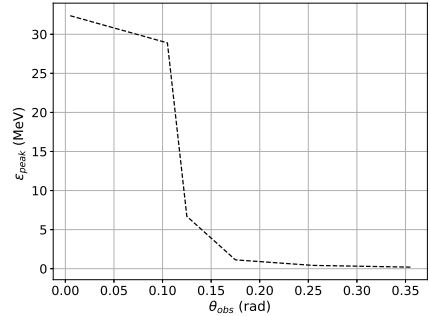}
\caption{Variation of the spectral peak energy $\varepsilon_{peak}$ with observer's angle $\theta_{obs}$ for the parameters as in Figure \ref{aba:fig2}.}
\label{aba:fig4}
\end{center}
\end{figure}

Keeping the same parameters as above, we plot the variation of $\varepsilon_{peak}$ with observer's position $\theta_{obs}$ in Figure \ref{aba:fig4} and find that it varies from few 10 MeV to few KeV as the observing angle changes from 0 to 0.35 rad. The evolution of peak energies are followed due to relativistic kinematics. The photons are blueshifted for an on axis observer while a far off axis observer sees the photons to be having relatively less energies. This shifts the spectra to lower energies and subsequently low peak energies are observed. The obtained range of peak energies explains the typically observed range of the spectral peaks in GRB prompt phase spectra \cite{ghirlanda2011short}.

\begin{figure}
\begin{center}
\includegraphics[width=2.5in]{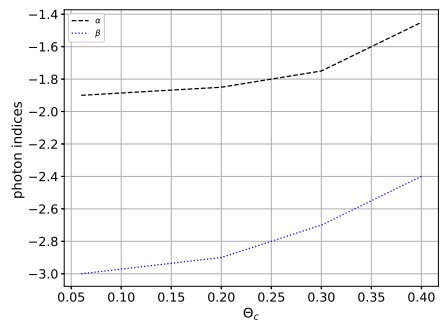}
\caption{Variation of photon indices $\alpha$ and $\beta$ with the cork temperature $\Theta_c$ for $\Theta_r=3$ and $\theta_{obs}=0.105$ rad.}
\label{aba:fig5}
\end{center}
\end{figure}

Finally, in Figure \ref{aba:fig5} we show that the photon indices $\alpha$ and $\beta$ are sensitive to the cork temperature $\Theta_c$. In the range $\Theta_c=1-10$, the magnitude of both indices decreases making the spectra harder. It is understandable as the spectrum becomes harder due to high energy electrons in cork at high temperatures.

\section{Conclusions}
In this proceedings, we have shown that the typical spectrum of a gamma ray prompt phase is consistent with the theoretical prediction of backscattering dominated picture. We show that the spectrum has negative $\alpha$ and steeper $\beta$. The ranges obtained for both the indices is $\alpha\sim -1.9$ to $-1.1$ and $\beta \sim -3$ to $-2.4$. The spectral peak energies decrease as the observer is located farther from the axis or $\theta_{obs}$ is greater. In such as case, the peak energies are obtained to be in the range $\varepsilon_{peak}=$few 10 KeV to few 10 MeV. All the obtained ranges of these parameters, $\alpha$, $\beta$ and $\varepsilon_{peak}$ are within the observed ranges of GRB prompt phase observations \citep{preece2000batse, ghirlanda2011short, pe2015physics}.

\pagebreak

\section{Acknowledgments}
After recent demise of our collaborator late David Eichler, the completion of this work is a tribute to him.

\bibliographystyle{ws-procs961x669}
\bibliography{ws-pro-sample}

\end{document}